\documentclass[aps,showpacs,superscriptaddress,nofootinbib,floatfix,
11pt]{revtex4}
\pdfoutput = 1
\usepackage{graphicx}
\usepackage{color}
\usepackage{amsfonts,amssymb,latexsym,amsmath} 
\usepackage{bm}
\usepackage{array} 
\usepackage{epstopdf}
\usepackage{hyperref}

\newcommand{\Amp}{\mathcal{A}}
\newcommand{\order}[1]{\mathcal{O}\left(#1\right)}
\newcommand{\lam}{\Lambda_\text{QCD}}
\newcommand{\br}[1]{\mathcal{B}\left(#1\right)}
\newcommand{\al}{&}

\begin{document}
\title{ Enhanced breaking of heavy quark spin symmetry }

\author{Feng-Kun Guo}
\email{fkguo@hiskp.uni-bonn.de}
\affiliation{Helmholtz-Institut f\"ur Strahlen- und
             Kernphysik and Bethe Center for Theoretical Physics, \\
             Universit\"at Bonn,  D--53115 Bonn, Germany}

\author{Ulf-G. Mei{\ss}ner}
\email{meissner@hiskp.uni-bonn.de}
\affiliation{Helmholtz-Institut f\"ur Strahlen- und
             Kernphysik and Bethe Center for Theoretical Physics, \\
             Universit\"at Bonn,  D--53115 Bonn, Germany}
\affiliation{Institute for Advanced Simulation, Institut f\"{u}r Kernphysik
             and J\"ulich Center for Hadron Physics, \\
             Forschungszentrum J\"{u}lich, D--52425 J\"{u}lich, Germany}

\author{Cheng-Ping Shen}
\email{shencp@ihep.ac.cn}
\affiliation{School of Physics and Nuclear Energy Engineering,
             Beihang University, Beijing 100191, China }

\begin{abstract}
\noindent
Heavy quark spin symmetry  is useful to make predictions on ratios of
decay or production rates of systems involving heavy quarks. The breaking of spin
symmetry is generally of  the order of $\order{\lam/m_Q}$, with $\lam$ the scale
of QCD and $m_Q$ the heavy quark mass.  In this paper, we
will show that a small $S$- and $D$-wave mixing in the wave function of the
heavy quarkonium could induce a large breaking in the ratios of partial decay
widths. As an example, we consider the decays of the $\Upsilon(10860)$ into the
$\chi_{bJ}\omega\, (J=0,1,2)$, which were recently measured by the Belle
Collaboration. These decays exhibit a huge breaking of the spin symmetry 
relation were
the $\Upsilon(10860)$ a pure $5S$ bottomonium state. We propose that this could
be a consequence of a mixing of the $S$-wave and $D$-wave components in the
$\Upsilon(10860)$. Prediction on the ratio
$\Gamma(\Upsilon(10860)\to\chi_{b0}\omega)/
\Gamma(\Upsilon(10860)\to\chi_{b2}\omega)$ is presented assuming that the decay
of the $D$-wave component is dominated by the coupled-channel effects.

\end{abstract}



\maketitle

\newpage


A heavy quarkonium is a system consisting of a heavy quark
and a heavy antiquark.
The ground states and low-lying excited states below the open-flavor
thresholds were well described in terms of potential quark models, e.g., the
Godfrey--Isgur quark model~\cite{Godfrey:1985xj}, while the higher excited
states are more complicated. The complexity comes from, e.g., the nearby
strongly coupled thresholds, the existence of many new quarkonium-like states
discovered in the last decade and so on.
Because the heavy quark mass $m_Q$ is much larger than the
scale of quantum
chromodynamics (QCD), $\lam$, the amplitude of changing the spin
orientation of a heavy quark by interacting with soft gluons is small,
suppressed by $\order{\lam/m_Q}$ relative to the spin-conserving
case~\cite{Caswell:1985ui}. The resulting heavy quark spin symmetry
(HQSS)~\cite{Isgur:1989vq} can lead to important observable
consequences. On the one hand, heavy quarkonium states are organized into spin
multiplets; on the other hand, the decay or production rate involving
one heavy quarkonium can often be related to the one of its spin partners in the
leading approximation. Breaking of HQSS is typically of the order of
$\order{\lam/m_Q}$ or even higher. In this paper, we will argue that the
HQSS breaking could be much larger in certain processes.
To be specific, we will show that a small mixing of $S$- and $D$-wave heavy
quarkonia could result in a
significant breaking of the spin symmetry relations when the decay amplitude of the $D$-wave component is enhanced. As an example,
we will calculate the processes $\Upsilon(10860)\to\chi_{bJ}\omega\,
(J=0,1,2)$. Measurements for these transitions were done by the Belle
Collaboration very recently, and the results for
the branching fractions are~\cite{He:2014sqj}
\begin{eqnarray}
    \br{\Upsilon(10860)\to\chi_{b0}\omega} &<& 3.9\times 10^{-3}, \nonumber\\
    \br{\Upsilon(10860)\to\chi_{b1}\omega} &=&
    (1.57\pm0.22_\text{stat.}\pm0.21_\text{sys.})\times 10^{-3}, \nonumber\\
    \br{\Upsilon(10860)\to\chi_{b2}\omega} &=&
    (0.60\pm0.23_\text{stat.}\pm0.15_\text{sys.})\times 10^{-3}.
    \label{eq:belle}
\end{eqnarray}
One sees that the branching fraction for the $\chi_{b1}\omega$ mode is larger
than that for the $\chi_{b2}\omega$.  Comparing
the HQSS prediction on the ratio $\br{\Upsilon(5S)\to\chi_{b1}\omega}/
\br{\Upsilon(5S)\to\chi_{b2}\omega}=0.63$
assuming the $\Upsilon(10860)$ to be the $5S$ bottomonium state,
see Eq.~\eqref{eq:Sratio_PS} below, with the observed value $2.62\pm1.30$, the
breaking is more than 100\%. This  is a very large spin symmetry
breaking.
As we will show later, a small mixture of a $D$-wave $\bar b b$ component
in the $\Upsilon(10860)$ is able to cause the
ratios of $\Gamma(\Upsilon(10860)\to\chi_{bJ}\omega)$ to be very different from the
spin symmetry relations as observed.

Consequences of HQSS can be easily analyzed using heavy meson effective
field theory (for a review, see Ref.~\cite{Casalbuoni:1996pg}).
Let us take the transitions from a vector heavy quarkonium into the
$\chi_J\omega$ as an example, where $\chi_J$ is a $P$-wave heavy quarkonium
with quantum numbers $J^{PC}=J^{++}$.
Here we will use the two-component notation in Ref.~\cite{Hu:2005gf} which is
convenient for nonrelativistic processes with negligible recoil effect.
The fields for the $S$-wave, $P$-wave and $D$-wave heavy quarkonium states are
denoted by $J$, $\chi^i$ and $J^{ij}$, respectively, which
are
$J = \vec{ \psi}\cdot \vec{\sigma}$,
$\chi^i = \sigma^j \left( \delta^{ij} \chi_0/\sqrt{3}   -
  \epsilon^{ijk} \chi_1^k /\sqrt{2} -\chi_2^{ij} \right)$,
$ J^{ij} = \frac{3}{2\sqrt{15} } \left( \psi_D^i \sigma^j +
\psi_D^j\sigma^i \right) - \frac1{\sqrt{15} }\delta^{ij}\vec{\psi}_D\cdot \vec \sigma
$~\cite{Casalbuoni:1996pg,Fleming:2008yn,Guo:2010ak,Margaryan:2013tta},
where $\vec\sigma$ are the Pauli matrices, and $\psi$, $\chi_J$ and $\psi_D$
annihilate the $S$-, $P$- and $D$-wave heavy quarkonia, respectively. The
states included in the above expressions have other spin partners which can be
included as well, however, only the fields relevant for our discussion are
shown.

Since the heavy quarkonia can be treated nonrelativistically, an expansion
over low momenta can be done. To leading order of such an expansion, the
Lagrangian for the decays of an $S$-wave or a $D$-wave heavy quarkonium into
$\chi_J\omega$ reads
\begin{equation}
    \mathcal{L}_{\chi\omega} = \frac{c_S}{2} \left\langle \chi^{i\,\dag} J
\right\rangle \omega^i +
\frac{c_D}{4} \left(  \left\langle \chi^{i\,\dag} J^{ij}
\right\rangle \omega^j +  \left\langle \chi^{j\,\dag} J^{ij}
\right\rangle \omega^i \right),
\end{equation}
where $\langle\,\rangle$ denotes the trace over the spinor space.
With this Lagrangian,  one is ready to obtain the ratios of decay widths of
an excited $ S $-wave heavy quarkonium into the $\chi_J\omega$ when
the difference in phase space is neglected
\begin{eqnarray}
\Gamma(\psi\to\chi_0\omega) : \Gamma(\psi\to\chi_1\omega) :
\Gamma(\psi\to\chi_2\omega)  = 1:3:5.
\label{eq:Sratio_noPS}
\end{eqnarray}
The ratios are completely different if the initial state is a $D$-wave heavy
quarkonium. In this case, one obtains
\begin{eqnarray}
 \Gamma(\psi_D\to\chi_0\omega) : \Gamma(\psi_D\to\chi_1\omega) :
\Gamma(\psi_D\to\chi_2\omega)
 = 20:15:1,
\label{eq:Dratio_noPS}
\end{eqnarray}
Therefore, the ratios of the decay widths of
an excited heavy quarkonium into the $\chi_{J}\omega$ can be used to probe the
spin structure of the initial state.

Replacing the $\omega$ by a photon, the above analysis still applies if we
change the widths on the left side of Eqs.~\eqref{eq:Sratio_noPS} and
\eqref{eq:Dratio_noPS} by $\Gamma/E_\gamma^3$ with $E_\gamma$ the photon
energy in the rest frame of the initial state. The factor of the photon energy is
required by gauge symmetry. As was shown
long time ago in Ref.~\cite{Cho:1994ih}, the spin symmetry
relations for the radiative transitions are generally in a quite good
agreement with the experimental data, and the breaking of
the spin symmetry relations is at the order of
$\order{\lam/m_Q}$.

However, HQSS breaking for near-threshold vector quarkonium states could be
enhanced due to the coupling to heavy meson pairs in a
$P$-wave~\cite{Voloshin:2012dk}. In the following, we will explore a different
mechanism, and show that a small $S$-$D$ mixing \footnote{In our
case of the decays $\Upsilon(10860)\to\chi_{bJ}\omega$, as will be shown later a
mixing angle of $\order{\lam^2/m_b^2}\sim 1^\circ$ is not sufficient. However,
if the mixing angle can be enhanced to around $5^\circ$, which is
still small, or larger, the huge HQSS breaking observed by the Belle
Collaboration can be explained by the mechanism proposed here.
Phenomenologically, the mixing angle for the $\Upsilon(10860)$ could be larger
than $20^\circ$~\cite{Badalian:2009bu}. } could result in a significant spin
symmetry breaking
 if the decays of the $D$-wave component are enhanced by, for
instance, coupled-channel effect as will be considered in the following.

Let us take the decays of the $\Upsilon(10860)$ into the $\chi_{bJ}\omega$ as
a specific example. The $\Upsilon(10860)$ is often considered as the $5S$
vector bottomonium. It was argued that the HQSS breaking in the
$\Upsilon(10860)$ decays into open-bottom mesons could be as large as
10\% to 20\%~\cite{Voloshin:2013ez} (see also discussions in
Ref.~\cite{Mehen:2013mva}). It is thus reasonable to assume that the wave
function of the $\Upsilon(10860)$ contains a small mixture of a $D$-wave
component, $\Upsilon_D$. The decay amplitude  can be written as
\begin{equation}
    \Amp(\Upsilon(10860)\to \chi_{bJ}\omega ) = \cos\theta\,\Amp_S +
\sin\theta\,\Amp_D,
\label{eq:sdmixing}
\end{equation}
where $\theta$ is the mixing angle, and $\Amp_{S}$ and $\Amp_D$ are the decay
amplitudes from the the $S$-wave and $D$-wave components, respectively. One
sees from Eqs.~\eqref{eq:Sratio_noPS} and \eqref{eq:Dratio_noPS} that the
ratios of the partial widths of the $S$-wave and $D$-wave components are
distinct. When the phase space is taken into account, the corresponding ratios
for the $\Upsilon(10860)$ decays in question are
\begin{equation}
    \Gamma_0^S:\Gamma_1^S:\Gamma_2^S=1:2.8:4.4,
    \label{eq:Sratio_PS}
\end{equation}
and
\begin{equation}
    \Gamma_0^D:\Gamma_1^D:\Gamma_2^D=22.9:15.8:1
    \label{eq:Dratio_ps}
\end{equation}
respectively, where $\Gamma_J$ represents
$\Gamma(\Upsilon(10860)\to\chi_{bJ}\omega)$, and the index $S(D)$ means that
only the $S(D)$-wave component is considered.

Thus, if there is a mechanism to enhance the decay amplitude of the
$D$-wave component relative to one of the $S$-wave component, a relatively
small $D$-wave admixture can induce a sizable breaking of HQSS. In the
following, we will assume that the decay width from the $S$-wave component is
very small, and investigate the possibility of enhancing HQSS breaking due
to such a mixing.

\begin{figure}
    \centering
    \includegraphics[width=0.5\textwidth]{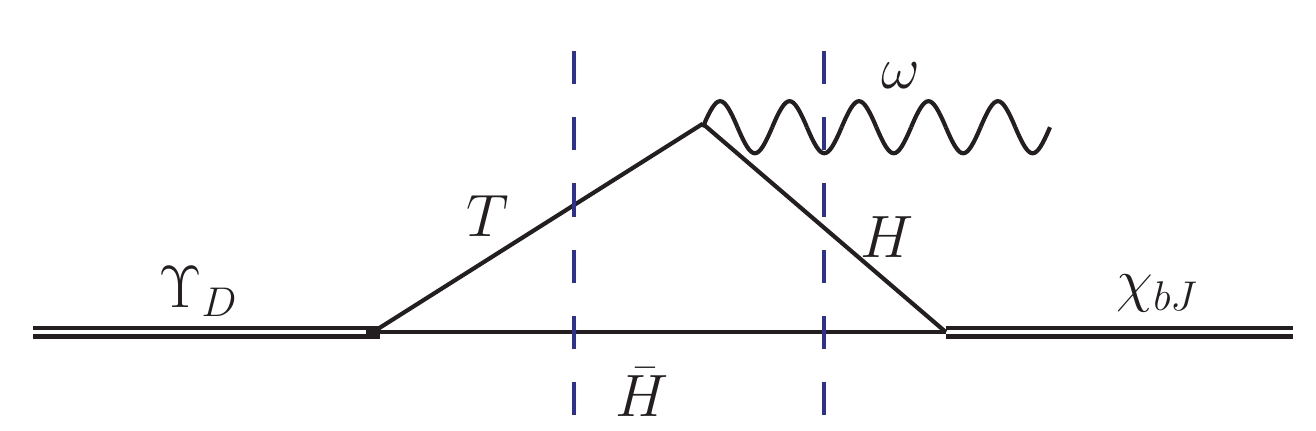}
    \caption{The dominant decay mechanism for the $D$-wave component of the
$\Upsilon(10860)$ into the $\chi_{bJ}\omega $. Here, $\Upsilon_D$ denotes the
$D$-wave component, and $T$ and $H(\bar H)$ represent the bottom mesons with
$s_\ell^P=\frac32^+$ and $\frac12^-$, respectively. The charge conjugated
diagram is not shown but taken into account in the calculations.
The vertical dashed lines indicate the two cuts operative in the process. }
\label{fig:triangle}
\end{figure}

As analyzed in details in Ref.~\cite{Guo:2010ak} for the transitions between two
charmonium states with the emission of a pion or $\eta$-meson, some decay
processes could be dominated by coupled-channel effects due to the coupling to
the intermediate virtual heavy and anti-heavy mesons. Especially, the
coupled-channel effect is the most important when both the vertices involving
heavy quarkonia are in an $S$-wave. The mass of the $\Upsilon(10860)$ is only
about 120~MeV below the threshold of the $B_1(5721)\bar B$. Thus, the decays of
the $D$-wave component of the $\Upsilon(10860) $ could be dominated by meson
loops as shown in Fig.~\ref{fig:triangle}. This is analogous to the radiative
decays of the $D$-wave charmonia into the $X(3872)$~\cite{Guo:2013zbw}. The
hypothesis is based on a nonrelativistic power counting in terms of the velocity
of the intermediate heavy mesons, denoted by $v$.
Because both the initial and final heavy quarkonia are not far from the
thresholds of the coupled heavy mesons, the intermediate heavy mesons are
nonrelativistic with a velocity $v\ll1$.  For the diagram shown in
Fig.~\ref{fig:triangle}, all three vertices are $S$-wave, and thus the loop
amplitude is of the order $\mathcal{O}(v^5/(v^2)^3) = \mathcal{O}(v^{-1})$,
where $v^5$ and $(v^2)^3$ account for the measure of the loop integral and three
nonrelativistic propagators, respectively.
Since the both the initial and final bottomonia are not far away from the
threshold of the bottom meson pair, two unitary cuts are operative in this
diagram, shown by the dashed vertical lines in Fig.~\ref{fig:triangle}. Each cut
corresponds to a momentum, and therefore a velocity. As discussed in
Appendix~\ref{app:loop}, the velocity in the power counting corresponds to the
average of the two velocities.
\begin{figure}
\centering
  \includegraphics[height=5.3cm]{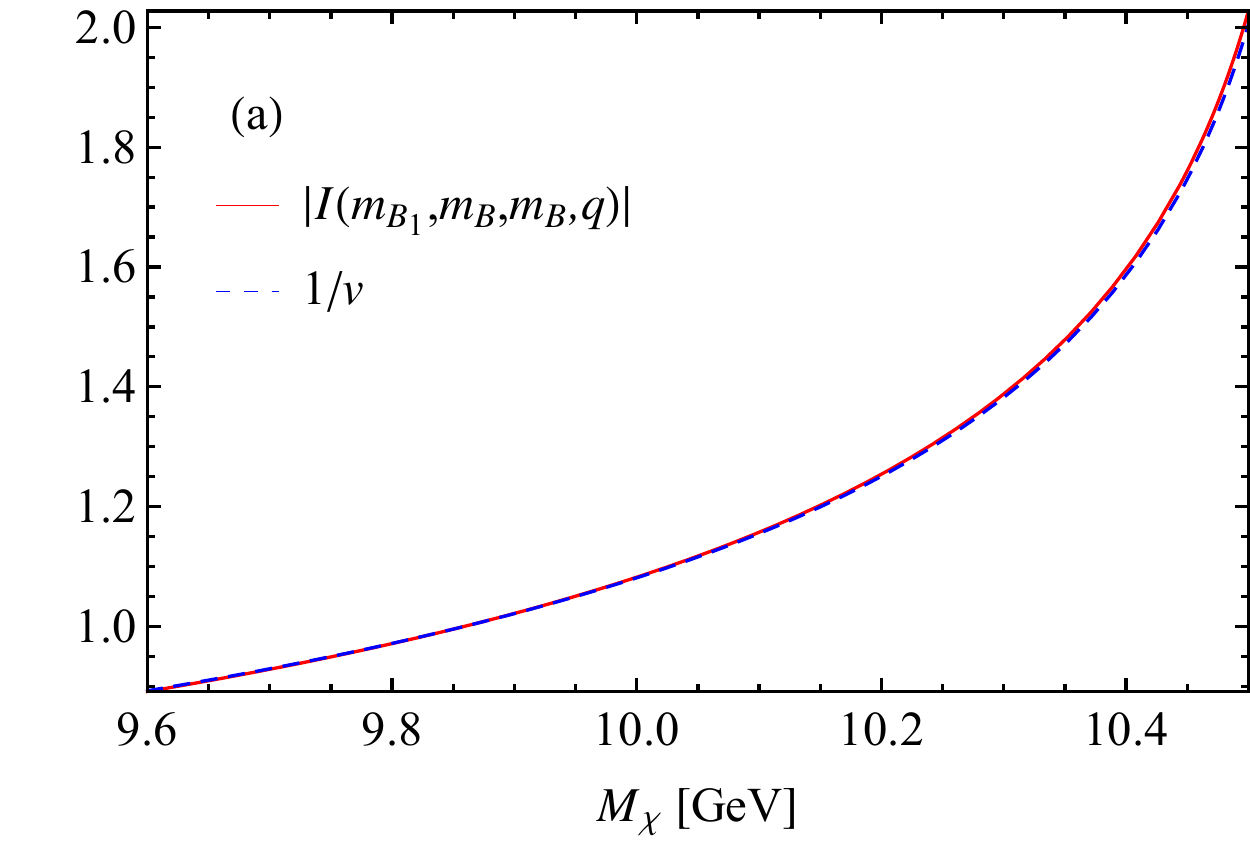}\hfill
  \includegraphics[height=5.3cm]{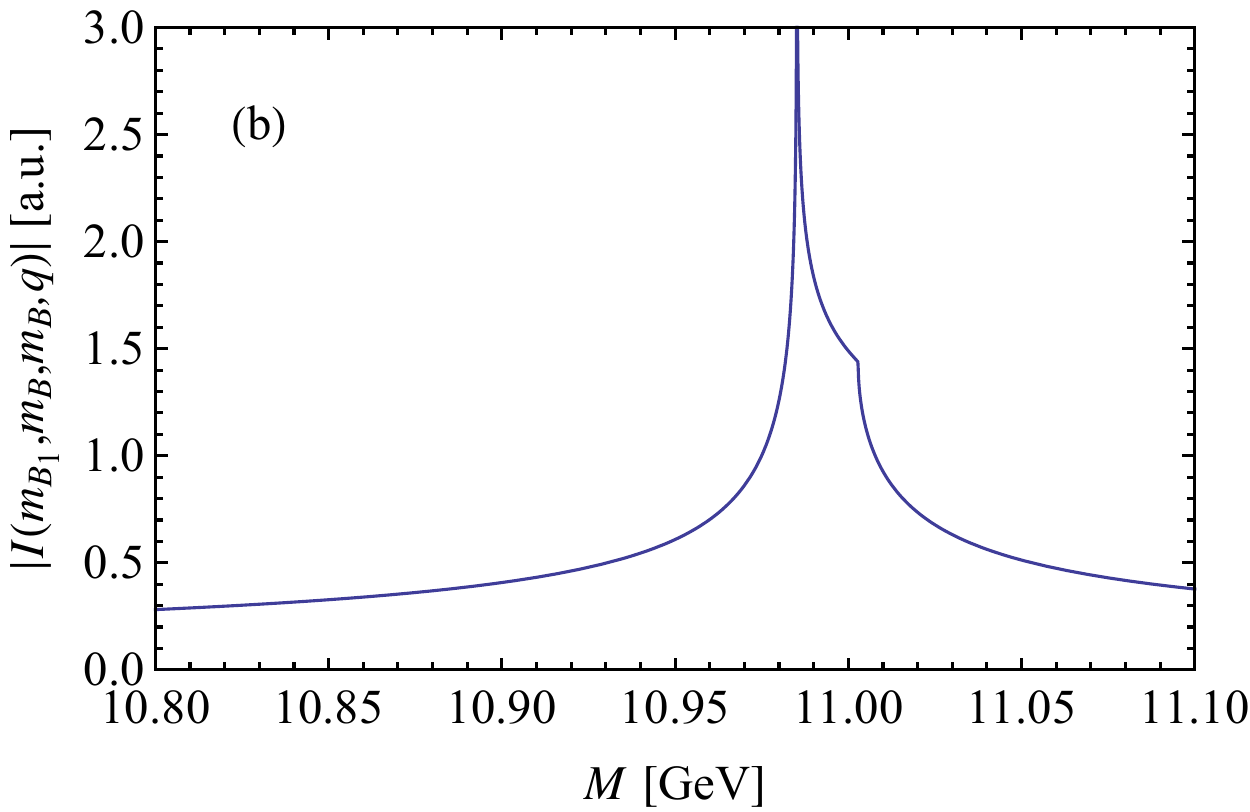}
  \caption{ (a) Illustration of the $v^{-1}$ scaling of the scalar
  three-point loop function. The solid curve represents
  $|I(m_{B_1},m_B,m_B,q)|$, see Eq.~\eqref{eq:nrloop}, with
  $M=M_{\Upsilon(10860)}$, and the dashed curve gives the inverse of the
  averaged velocity defined as $(v_1+v_2)/2$ with $v_1=\sqrt{ 2 \mu_1(
  m_{B_1}+m_B -M_{\Upsilon(10860)} )}/\bar m_1$, where $\mu_1$ and $\bar m_1$
  are the reduced and averaged masses of the $B_1$ and $B$, respectively, and
  $v_2=\sqrt{ ( 2 m_B - M_\chi )/m_B }$.
  For comparison, both the loop function and $1/v$ are normalized at
  $M_\chi=M_{\chi_{b0}}$. (b) Dependence of $|I(m_{B_1},m_B,m_B,q)|$ evaluated
  at $M_\chi=2M_B$ on the mass of the initial state $M$. }
  \label{fig:pc}
\end{figure}
 This can be seen from a comparison of the scalar
three-point loop function and the inverse of the averaged velocity as shown in
Fig.~\ref{fig:pc}(a). Notice that although the loop function scales as $v^{-1}$,
it does not diverge even when both masses of the initial and final heavy
quarkonium states are located at the corresponding thresholds. In
Fig.~\ref{fig:pc}(b), we show $|I(m_{B_1},m_B,m_B,q)|$ evaluated at
$M_\chi=2M_B$ as a function of $M$. One sees that at threshold $M=m_{B_1}+m_B$,
there is a cusp which is due to square-root singularity at the
threshold; the sharp peak below the cusp is due to the Landau singularity
discussed in Appendix~\ref{app:loop}. For the processes in question, we have 
the averaged
velocity $v\approx 0.26$.
Therefore, the negative power of the small velocity provides an enhancement
to the coupled-channel amplitudes. Thus, it is reasonable to assume
that the decays of the $D$-wave component into the $\chi_{bJ}\omega$ are
dominated by the loop diagrams as shown in Fig.~\ref{fig:triangle}, and the
partial widths are not small. For more discussion of the power counting, we
refer to
Refs.~\cite{Guo:2009wr,Guo:2010ak,Guo:2012tg,Cleven:2011gp,Cleven:2013sq}.
Next, we will perform an explicit calculation of the coupled-channel effect
based on the mechanism shown in Fig.~\ref{fig:triangle}.

In the two-component notation, the fields for the $S$-wave
($s_\ell^P=\frac12^-$) and $P$-wave
($s_\ell=\frac32^+$) heavy mesons read
$    H_a = \vec{V}_a\cdot \vec{\sigma} + P_a$, and $
    T_a^i = P_{2a}^{ij} \sigma^j + \sqrt{2/3}\, P_{1a}^i + i
\sqrt{1/6}\, \epsilon_{ijk} P_{1a}^j \sigma^k$,
where $P_a$ and $V_a$ annihilate the pseudoscalar and vector heavy mesons,
respectively, with $a=u,d$ labeling the light flavors, and $P_{1a}$ and
$P_{2a}$ annihilate the axial and tensor heavy mesons, respectively. The fields
annihilating their anti-particles are
  $\bar H_a = -\vec{\bar V}_a\cdot \vec{\sigma} + \bar P_a$,
  $\bar T_a^i = -\bar P_{2a}^{ij} \sigma^j + \sqrt{2/3}\, \bar
  P_{1a}^i - i \sqrt{1/6}\, \epsilon_{ijk} \bar P_{1a}^j \sigma^k$.
The properties of these fields under symmetry transformations can be found in
Refs.~\cite{Fleming:2008yn,Guo:2013zbw}.

The Lagrangian, which is invariant under transformations of parity, charge
conjugation, HQSS and Galilean invariance, for the coupling of the $P$-wave and
$D$-wave heavy quarkonia to the $s_\ell=\frac12^-$ and $s_\ell=\frac32^+$ heavy
mesons to leading order of the nonrelativistic expansion can be written
as~\cite{Colangelo:2003sa,Fleming:2008yn,Guo:2013zbw}
\begin{eqnarray}
  \label{eq:lagPD}
  \mathcal{L}_{PD} = \frac{g_4}{2} \left \langle
\left( \bar{T}^{j\,\dag}_a\, \sigma^i H^\dag_a - \bar{H}^\dag_a \,\sigma^i
T^{j\,\dag}_a \right) J^{ij} \right \rangle
 + \frac{g_1}{2} \left\langle \chi^{i\dag} H_a\sigma^i \bar
H_a \right\rangle + \text{H.c.}\,.
\end{eqnarray}
The $S$-wave coupling of the $\omega$-meson to the $S$-wave and $P$-wave heavy
mesons can be described by
\begin{equation}
    \label{eq:lagomega}
    \mathcal{L}_{\omega} = \frac{c_\omega}{2} \left\langle H_a^\dag T_a^i -
\bar H_a^\dag \bar T_a^i \right\rangle \omega^i + \text{H.c.}\, ,
\end{equation}
where isospin symmetry is assumed.

\begin{table*}[tb]
\begin{center}
\begin{ruledtabular}
   \caption{ Heavy meson loops contributing to the decays of
the vector $D$-wave bottomonium into the $\chi_{bJ}\omega$. Here the charge
conjugated ones are not listed but considered in the calculation. }
\label{tab:loops}
   \begin{tabular}{| l | c c c|}
      Processes &  $\Upsilon_D\to\chi_{b0}\omega $  &
$\Upsilon_D\to\chi_{b1}\omega $ & $\Upsilon_D\to\chi_{b2}\omega $ \\ \hline
      Loops & $[B_1\bar B B], [B_1\bar B^* B^*], [B_2\bar B^* B^*]$
			&  $[B_1\bar B B^*], [B_1\bar B^* B]$
			&  $[B_1\bar B^* B^*], [B_2\bar B^* B^*]$
            \\
   \end{tabular}
   \end{ruledtabular}
\end{center}
\end{table*}
Denoting the diagram shown in Fig.~\ref{fig:triangle} by $[T\bar H
H]$, the loops contributing to the processes $\Upsilon_D\to \chi_{bJ}\omega$
are listed in Tab.~\ref{tab:loops}.
Using the Lagrangians given in Eqs.~\eqref{eq:lagPD} and \eqref{eq:lagomega},
the decay amplitudes can be easily obtained, and the explicit
expressions are given in Appendix~\ref{app:loop}.
It is interesting to notice that if we take the same mass for the heavy mesons
in the same spin multiplet, the spin symmetry relations are kept even if
coupled channels are considered, that is, one would get the same ratios
$20:15:1$ for
$\left|\mathcal{A}_{\Upsilon_D\to\chi_{bJ}\omega}^\text{loop}\right|^2$ as the
ones in Eq.~\eqref{eq:Dratio_noPS}.
This can be understood because the  Lagrangians respect spin symmetry, and
if we use degenerate masses, there will be no source for symmetry
breaking.~\footnote{This provides a simple method to calculate the HQSS
relations for partial decay widths of processes involving hadronic molecules
of a pair of heavy mesons, and the results in, e.g., Ref.~\cite{Ma:2014ofa}
calculated using 6-$j$ and 9-$j$ symbols can be checked in this way. Without
taking into account the phase space factors which include $E_\gamma^3$ for the
radiative decays and setting mesons in the same spin multiplet to be
degenerate, the ratios for the decay widths, $\Gamma_J^\text{hm}$, of a
$1^{--}$ bottomonium-like hadronic molecule into $\chi_{bJ}\omega/\gamma$ are
as follows: $\Gamma_0^\text{hm}:\Gamma_1^\text{hm}:\Gamma_2^\text{hm}=3:1:0$
(for $B_1\bar B$), $1:12:5$ (for $B_1\bar B^*$), and $5:0:1$  (for $B_2\bar
B^*$). }
When the physical masses for all the mesons are used, and the phase space
difference is taken into account, the loop amplitudes will result in ratios
slightly different from Eq.~\eqref{eq:Dratio_ps}
\begin{equation}
    \Gamma_0^\text{loop}:\Gamma_1^\text{loop}:\Gamma_2^\text{loop}=24.4:16.7:1.
\label{eq:ratio_loop}
\end{equation}
One sees that the decays into the $\chi_{b0}\omega$ and $\chi_{b1}\omega$ are
more enhanced than that into the $\chi_{b2}\omega$. The reason is that the
$\Upsilon(10860)$ mass is closer to the $B_1\bar B$ threshold than to the
$B_1\bar B^*$ one, cf. Table~\ref{tab:loops}. If we put the initial state
at the mass of the $\Upsilon(11020)$, the heaviest known bottomonium, the
ratios will be even larger, $27.5:18.4:1$.

\begin{figure*}[tb]
    \centering
    \includegraphics[width=0.496\textwidth]{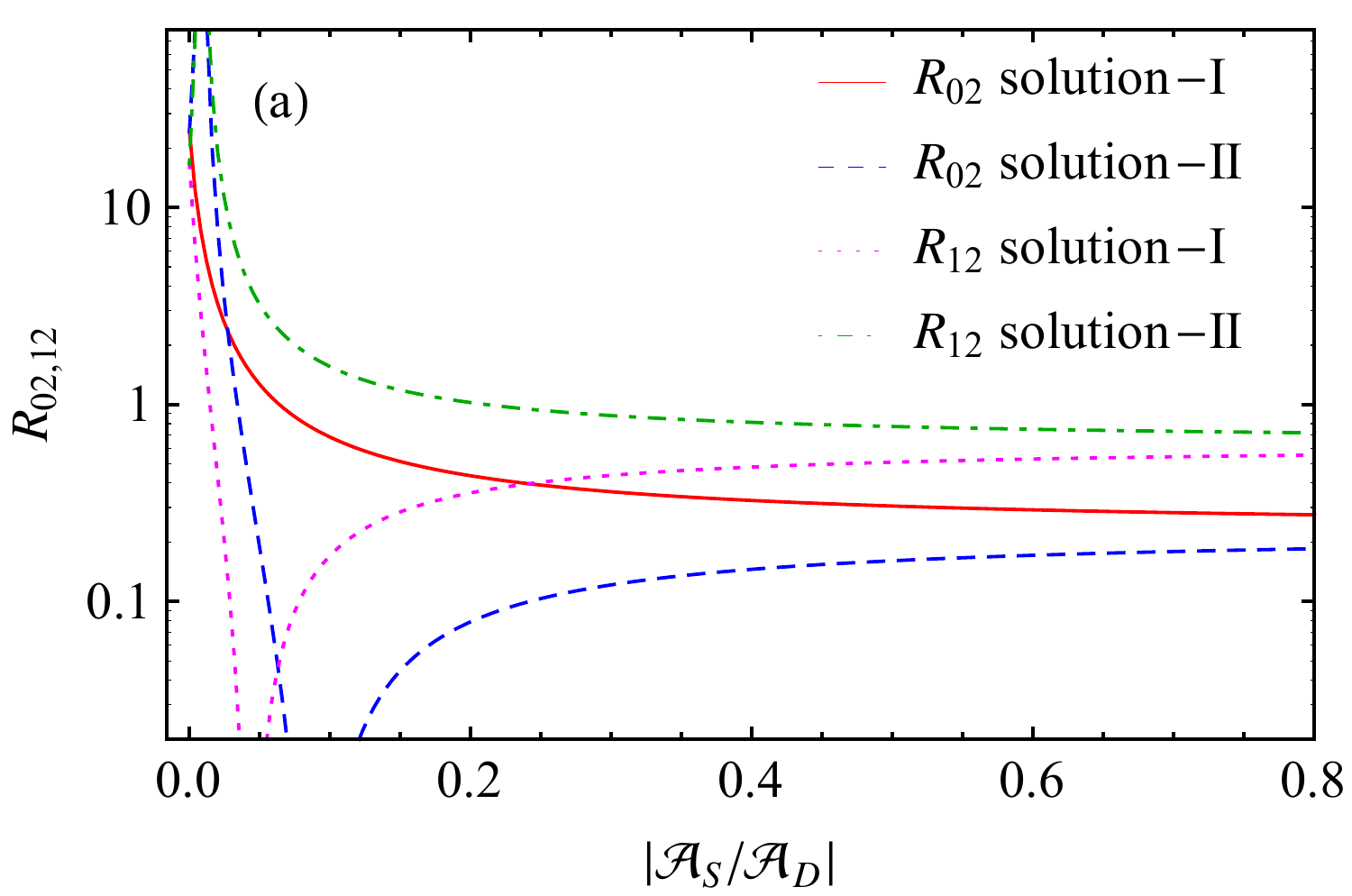}\hfill
    \includegraphics[width=0.496\textwidth]{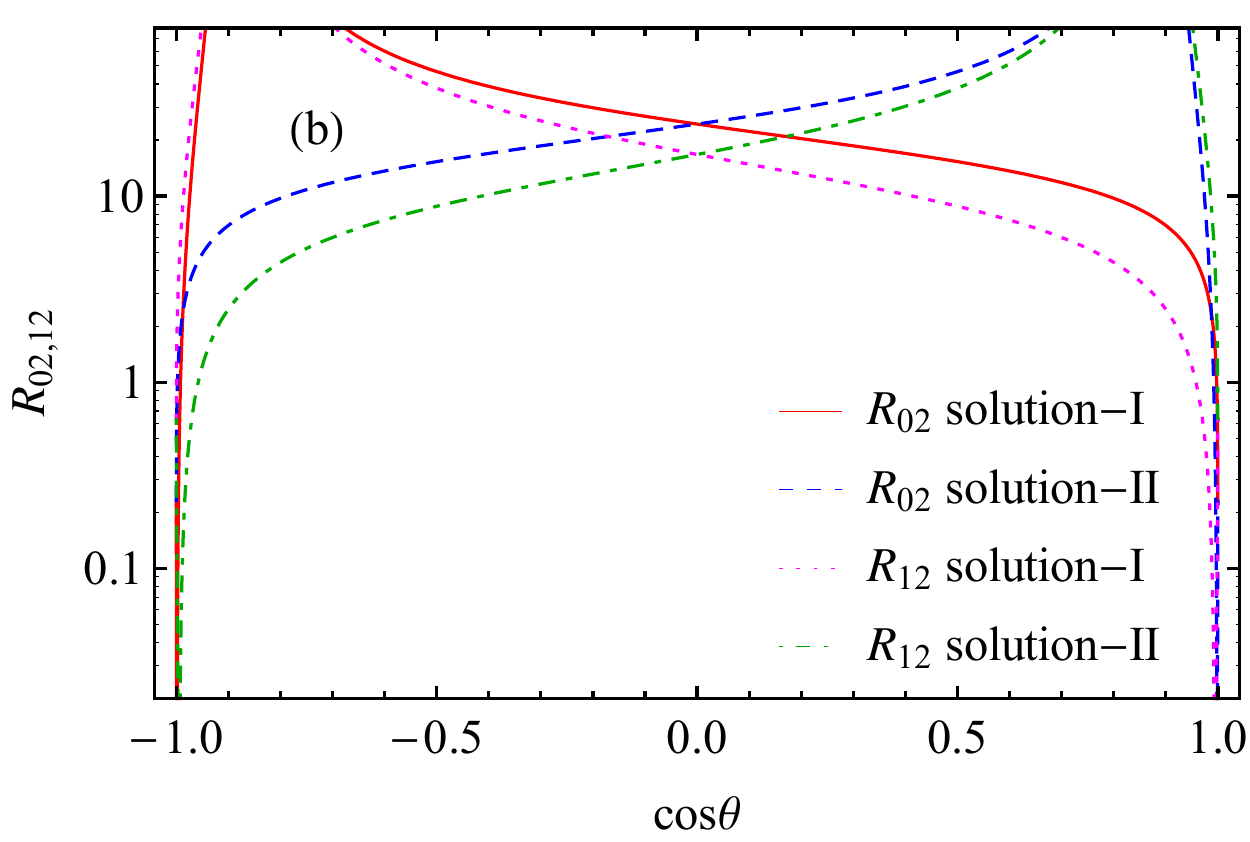}
    \caption{Dependence of the ratios $R_{02}$ and $R_{12}$ defined in
    Eq.~\eqref{eq:r0212} on $|\Amp_S/\Amp_D|$ for $\theta=5^\circ$ (a),
and on $\cos\theta$ for $|\Amp_S/\Amp_D|=0.05$ (b).
}
\label{fig:dependence}
\end{figure*}

With the above preparation, we can now show quantitatively how a significant
HQSS breaking effect can be obtained from a small $S$-$D$ mixing.
In the following, we will assume that the decays of the $D$-wave component into
the $\chi_{bJ}\omega$ are saturated by the triangle diagrams as discussed above.
Because the $S$-$D$ mixing is due to the tensor force between the heavy quark
and antiquark, it is of $\order{\lam^2/m_b^2}$, which corresponds to the mixing
angle $\lesssim1^\circ$ if $\lam$ is taken to be of the order of a few hundreds
MeV. However, as pointed out in Ref.~\cite{Badalian:2009bu}, for highly excited
bottomonia, the mass difference between the $(n+1)S$ and the $nD$ states is
small so that the mixing could be much larger. The phenomenological value for
the $\Upsilon(4S)$-$\Upsilon(3D)$ mixing angle extracted from the dielectron
width is as large as $27^\circ\pm4^\circ$, and the $5S$-$4D$ mixing angle is of
a similar size~\cite{Badalian:2009bu}.
Indeed, if we take $\theta=1^\circ$ and adjust the strength of the decay
amplitudes of the $S$-wave and $D$-wave components to get the central values of
$\Gamma(\Upsilon(10860)\to\chi_{b1}\omega)=(86\pm47)$~keV  and
$\Gamma(\Upsilon(10860)\to\chi_{b2}\omega)=(33\pm23)$~keV, one would get an
unreasonably large width for the $D$-wave component: two solutions are obtained
for $\Gamma_{D0}\equiv\Gamma(\Upsilon_D\to\chi_{b0}\omega)=604$~MeV or $75$~MeV.
These two values correspond to the ratio of the decay amplitude of the $S$-wave
component over that of the $D$-wave component $|\Amp_S/\Amp_D|=0.002$ and 0.011,
respectively.
These widths seem too large for an OZI-suppressed transition.
Increasing the angle to $5^\circ$, they become much more
reasonable---$\Gamma_{D0}=24$~MeV or $3$~MeV corresponding to
$|\Amp_S/\Amp_D|=0.008$ and 0.055, respectively. For a mixing angle as large as
$20^\circ$, one gets $\Gamma_{D0}=1.6$~MeV or 0.2~MeV corresponding to
$|\Amp_S/\Amp_D|=0.034$ and 0.23, respectively. In this regard, our explanation
of the large HQSS breaking in the partial decay widths in Eq.~\eqref{eq:belle}
requires the mixing angle between the $5S$ and the $4D$ states to be at least
around $5^\circ$.  One should also notice that 
according to the power counting of nonrelativistic QCD, the mixing is 
of the order $v_b^2\approx0.1$~\cite{Bodwin:1994jh}, where $v_b$ is the velocity 
of the bottom 
quark in bottomonium. In this sense, a mixing angle of $\mathcal{O}(10^\circ)$ 
is natural.
To be specific, let us take the mixing angle  $\theta=5^\circ$ for
instance, which corresponds to $\sin\theta=0.087$ and an $S$-wave dominance in
the wave function. In Fig.~\ref{fig:dependence}~(a), we show the dependence of
the ratios defined as
\begin{equation}
     R_{02} = \frac{\Gamma(\Upsilon(10860)\to\chi_{b0}\omega)}
     {\Gamma(\Upsilon(10860)\to\chi_{b2}\omega)}, \qquad
     R_{12} = \frac{\Gamma(\Upsilon(10860)\to\chi_{b1}\omega)}
      {\Gamma(\Upsilon(10860)\to\chi_{b2}\omega)} \label{eq:r0212}
\end{equation}
on $|\Amp_S/\Amp_D|$ for $\theta=5^\circ$. Because the interference between the
$S$-wave and $D$-wave components can be either constructive or destructive,
there are two possible solutions for each ratio. It is obvious that the
variation is dramatic at small values of $|\Amp_S/\Amp_D|$ due to interference.
This is because the contribution of the $D$-wave component is suppressed by the
small mixing angle, and the $S$-$D$ interference controls the results. Increasing
$|\Amp_S|$, the contribution from the $D$-wave
component diminishes, and the ratios approach those given in
Eq.~\eqref{eq:Sratio_PS}.
We thus
expect that  for small values of $| \Amp_S/\Amp_D |$ the ratios
would be very different from spin symmetry ones for the $\Upsilon(5S)$ given in Eq.~\eqref{eq:Sratio_PS}.
Figure~\ref{fig:dependence}~(b) shows the dependence on $\cos\theta$ for fixed
$|\Amp_S/\Amp_D|=0.05$.

\begin{figure}[tb]
   \centering
   \includegraphics[width=0.49\textwidth]{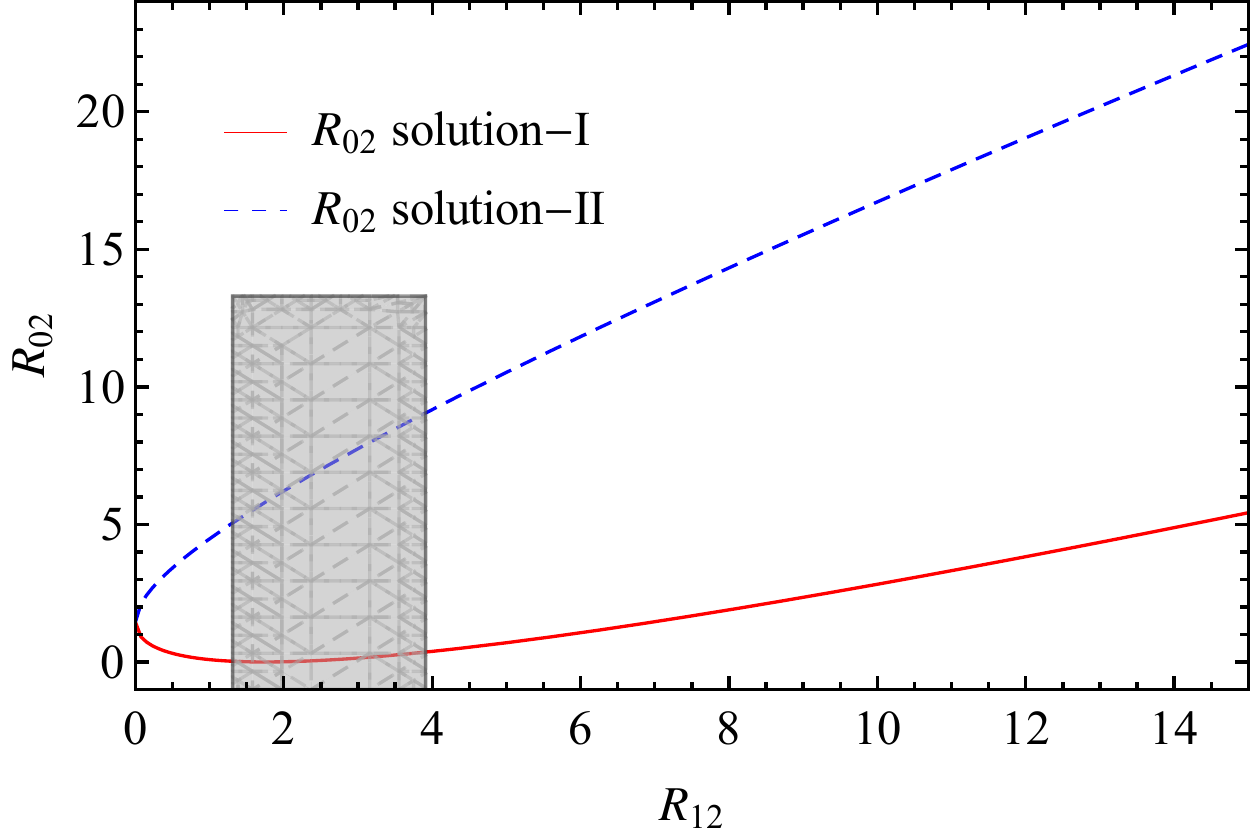}
   \caption{ Prediction of $R_{02}$ for a given value of $R_{12}$. The shaded
   area corresponds to the range reported by the Belle
   Collaboration~\cite{He:2014sqj}. }
\label{fig:r02}
\end{figure}

We want to emphasize that the mixing angle and $\Amp_S/\Amp_D$ always appear
together, and thus cannot be fixed from the measured branching fractions.
However, when one of the ratios $R_{02}$ or $R_{12}$ is measured, the other can
be predicted as shown in Fig.~\ref{fig:r02}, and the uncertainty should be of $\order{v}$.
In the figure, the Belle results $R_{12}=2.62\pm1.30$ and $R_{02}<13.3$ obtained
from Eq.~\eqref{eq:belle} are shown as the shaded area. Fixing $R_{12}$ to the
measured range, we predict two possible ranges for $R_{02}$,
\begin{equation}
    R_{02} = 7.1\pm2.1\pm1.8, \quad \text{or} \quad R_{02} = 0.19\pm0.17\pm0.05,
\end{equation}
where the first uncertainty is propagated from the measured uncertainty of
$R_{12}$, and the second one from $v=0.26$ is inherent in our nonrelativistic
framework. Both ranges are consistent with the Belle upper limit, and an
examination of the HQSS breaking mechanism proposed here urges an improved
measurement, especially for the $R_{02}$. This can be done at the future
super-$B$ factory. Similarly, we can make predictions for the decays
$\Upsilon(11020)\to\chi_{bJ}\omega$. The curves are similar with slightly larger values.

To summarize, we have discussed a new mechanism to produce a sizable breaking
of HQSS. We showed that a small $S$-$D$ mixing for
the vector heavy quarkonium could result in a much larger spin symmetry
breaking effect.  In order for this mechanism to work, the decays
of the $D$-wave component should be enhanced in comparison with that of the
$S$-wave one. As an example, we studied the decays
$\Upsilon(10860)\to\chi_{bJ}\omega$ in details. The decays of the $D$-wave
component of the $\Upsilon(10860)$ are assumed to be dominated by the
coupled-channel effects due to $S$-wave coupling to nearby thresholds of a
 $P$-wave and an $S$-wave heavy meson pair. It was found that a mixing angle of
 $\order{1^\circ}$ would result in a too large width for the $D$-wave component,
 and $\theta\gtrsim5^\circ$, i.e. $\sin\theta\gtrsim0.087$, is needed
 to explain the observed widths of the decays into the $\chi_{b1}\omega$
 and $\chi_{b2}\omega$. It is noticeable that
 a $\order{10\%}$ $D$-wave component, though needs an additional explanation
 for bottomonium states~\cite{Badalian:2009bu}, is sufficient to
 explain an HQSS breaking $\gtrsim100\%$ in the ratios of the partial decay 
widths.
In particular, when one of the ratios of branching fractions for the processes $\Upsilon(10860,11020)\to
\chi_{bJ}\omega$ is measured, the other can be predicted 
independent of the mixing angle.
Using the Belle measurement for $R_{12}$, two possible ranges of $R_{02}$ were 
predicted. The
prediction can be examined at the future super-$B$ factory. Such measurements
will be important to better understand the spin symmetry breaking
as well as the nature of the $\Upsilon(10860)$ and $\Upsilon(11020)$.

\medskip

\section*{Acknowledgments}

We are grateful to the referee for her/his valuable
comments on the first version of this paper. We would like to thank Bastian
Kubis for useful discussions. This work is supported in part by the DFG and the
NSFC through funds provided to the Sino--German CRC 110 ``Symmetries and the
Emergence of Structure in QCD'' (NSFC Grant No. 11261130311), by the EPOS 
network of the European Community
Research Infrastructure Integrating Activity ``Study of Strongly Interacting
Matter'' (HadronPhysics3, Grant No. 283286), by the NSFC (Grant No. 11165005), 
and by the
Fundamental Research Funds for the Central Universities (Grant No. 
YWF-14-WLXY-013) and CAS center for Excellence in Particle Physics (China).

\begin{appendix}

\renewcommand{\theequation}{\thesection.\arabic{equation}}

\section{Decay amplitudes and Landau singularities of the three-point loop
function}
\label{app:loop}
\setcounter{equation}{0}

The explicit expressions for the decay amplitudes for the $D$-wave component
through the $[T\bar HH]$ triangle diagrams are given by
\begin{eqnarray}
   \mathcal{A}_{\Upsilon_D\to\chi_{b0}\omega } ^\text{loop} \al=\al
-N\frac{2\sqrt{5}}{3}  g_1
g_4 c_{\omega }  \vec\varepsilon_{\Upsilon_D}\cdot \vec\varepsilon_\omega
\big[6 I\left(m_{B_1},m_{B},m_{B},{q} \right) +
I\left(m_{B_1},m_{B^*},m_{B^*},{q}\right) \nonumber\\
\al\al + I\left(m_{B_2},m_{B^*},m_{B^*}, {q} \right) \big],
 \nonumber\\
  \mathcal{A}_{\Upsilon_D\to\chi_{b1}\omega }^\text{loop} \al=\al 2N
\sqrt{\frac{10}{3}} g_1
g_4 c_{\omega } \epsilon_{ijk} \varepsilon ^i_{\Upsilon_D} \varepsilon
^j_{\omega}
\varepsilon^k_{\chi_{c1}}
   \left[ I\left(m_{B_1},m_{B},m_{B^*},{q} \right) +
   I\left(m_{B_1},m_{B^*},m_{B},{q} \right) \right], \nonumber\\
    \mathcal{A}_{\Upsilon_D\to\chi_{b2}\omega }^\text{loop} \al=\al
N\frac{2}{\sqrt{15}} g_1
g_4 c_{\omega } \varepsilon^{ij}_{\chi_{b2}} \varepsilon ^i_{\Upsilon_D}
\varepsilon
^j_\omega \left[  5 I\left(m_{B_1},m_{B^*},m_{B^*},{q}
\right) - I\left(m_{B_2},m_{B^*},m_{B^*},{q} \right) \right].
\label{eq:loopamp}
\end{eqnarray}
where  $N=\sqrt{M M_{\chi} } $, with $M$ and $M_\chi$ the masses of the initial
and final heavy particles, respectively, accounts for the nonrelativistic
normalization, $q$ is the magnitude of the three-momentum of the $\omega$ in the
rest-frame of the initial particle,
and $I(m_1,m_2,m_3,q)$ is the scalar three-point nonrelativistic loop integral,
the expression of which can be found in Refs.~\cite{Guo:2010ak,Guo:2013zbw}
\begin{eqnarray}
   \label{eq:nrloop}
   I(m_1,m_2,m_3,q)
   =\frac{\mu_{12}\mu_{23}}{16\pi\,m_1m_2m_3} \frac{1}{\sqrt{a}} \left[
   \arctan\frac{c'-c}{2\sqrt{a(c-i\epsilon)}} +
   \arctan\frac{2a+c-c'}{2\sqrt{a(c'-a-i\epsilon)}} \right],
\end{eqnarray}
with
\begin{equation}
     a = \left(\frac{\mu_{23}}{m_3} q\right)^2, \qquad
     c = 2\mu_{12}b_{12}, \qquad
     c'=2\mu_{23}b_{23}+ \frac{\mu_{23}}{m_3} q^2,
\end{equation}
where $\mu_{ij}=m_im_j/(m_i+m_j)$, $b_{12} = m_1+m_2-M$,
$b_{23}=m_2+m_3+E_\omega-M$, with $M$ the mass of the initial particle and
$E_\omega=(M^2-M_\chi^2+m_\omega^2)/(2M)$ the energy of the $\omega$-meson.

The meaning of the velocity $v$ in the power counting can be seen from expanding
the loop function around $a=0$~\cite{Guo:2012tg}
\begin{equation}
    I(m_1,m_2,m_3,q) = \frac{\mu_{12} \mu_{23}}{16\pi m_1m_2m_3}
    \frac{2}{\sqrt{c}+\sqrt{c'}} + \ldots,
\end{equation}
where only the leading order term is kept. Notice that the two square roots
inside the arctan functions in Eq.~\eqref{eq:nrloop} correspond to the two cuts
in Fig.~\ref{fig:triangle}. The one containing $\sqrt{c-i\epsilon}$ is
connected to the initial heavy quarkonium and cuts the intermediate states with masses
$m_1$ and $m_2$; the other, connected to the final heavy quarkonium, contains
$\sqrt{c'-a-i\epsilon}$ and cuts the intermediate states with masses $m_2$ and
$m_3$ and the light particle in the final state. It is thus clear that $v$ in
the power counting is the average of the two velocities defined through these
cuts.

Therefore, although the power counting of this scalar triangle loop is given by
$\order{v^{-1}}$, the loop function does not diverge even if $M=m_1+m_2$.
Indeed, the triangle loop integral has singularities in addition to the normal
thresholds which correspond to the branching points of the cuts. This has been
known for a long time~\cite{Landau:1959fi}, and such singularities are called
Landau singularities. Landau singularities for a given loop diagram are
determined by the solutions of the Landau equations. For the triangle diagram shown in
Fig.~\ref{fig:triangle}, the leading singularities are determined by the
following equation~\cite{Landau:1959fi}
\begin{equation}
    1 + 2\, y_{12}\, y_{23}\, y_{13} = y_{12}^2 + y_{23}^2 + y_{13}^2,
    \label{eq:Landau}
\end{equation}
where
\begin{equation}
    y_{ij} = \frac{m_i^2 + m_j^2 - p_{ij}^2}{2\,m_i\,m_j}.
\end{equation}
In our case, we have $p_{12}^2=M^2$, $p_{23}^2 = M_\chi^2$ and $p_{13}^2 =
m_\omega^2$.

\begin{figure}[tb]
   \centering
   \includegraphics[width=0.49\textwidth]{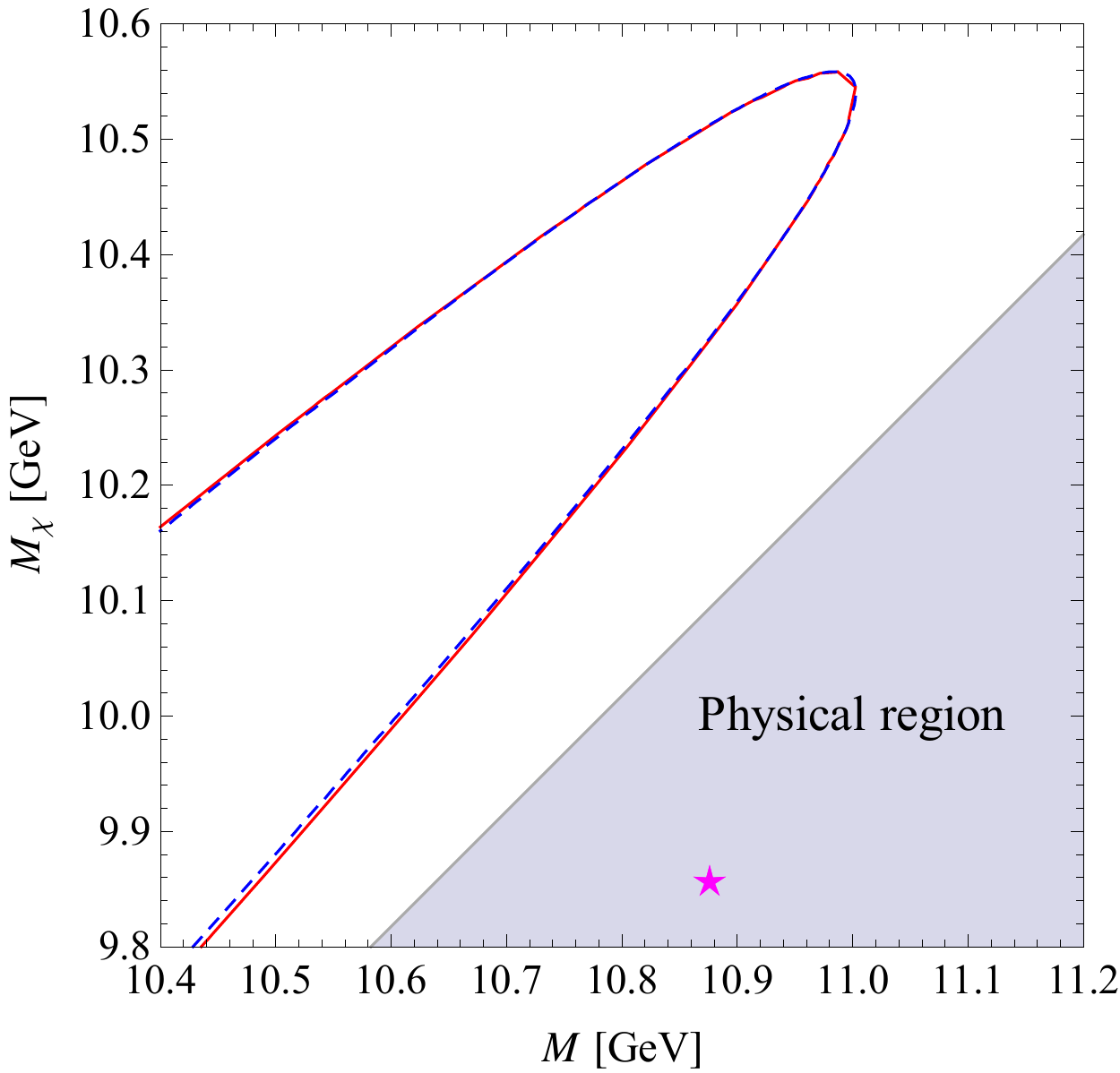}
   \caption{ The Landau singularity of the scalar triangle loop
   function for the intermediate mesons being $[B_1, \bar B,B]$. Here, $M$ and $M_\chi$ are
   the masses of the initial and final heavy quarkonia, and the light particle
   mass is $m_\omega$. The solid and dashed curves represent the trajectories
   for the solutions of the Landau equation, Eq.~\eqref{eq:Landau}, and the
   nonrelativistic equation, Eq.~\eqref{eq:NRsingularity}, respectively. The
   shaded area given by $M\geq M_\chi+m_\omega$ is the physically allowed
   region. The star marks the point with $M=M_{\Upsilon(10860)}$ and $M_\chi =
   M_{\chi_{b0}}$.}
\label{fig:singularity}
\end{figure}
As for the nonrelativistic triangle loop function in Eq.~\eqref{eq:nrloop},
the triangle singularity occurs when the arguments of the $\arctan$ functions take a value of $\pm i$. We find that the singularity equations from
both $\arctan$ functions are the same, which is
\begin{equation}
    (c'-c)^2 + 4 a c = 0.
    \label{eq:NRsingularity}
\end{equation}
Notice that this equation is of eighth order in the masses of the initial and
final heavy particles. Given a value of the initial mass, one gets eight
solutions for the mass of the final heavy particle $M_\chi$. However, since
Eq.~\eqref{eq:nrloop} is the expression for the nonrelativistic
three-point scalar loop integral, only those solutions of $M_\chi$ within the
vicinity of $m_2+m_3$ are valid. The solutions of interest of
Eq.~\eqref{eq:NRsingularity} are very close to those of Eq.~\eqref{eq:Landau}
as can be seen explicitly from Fig.~\ref{fig:singularity}.
They are not exactly the same because the Landau equations and thus
Eq.~\eqref{eq:Landau} are derived for relativistic propagators, while
Eq.~\eqref{eq:NRsingularity} is obtained from the nonrelativistic loop integral.
When the particle masses are real, Eq.~\eqref{eq:NRsingularity} can only be
satisfied when either $a$ or $c$ is non-positive, i.e. $q^2\leq0$ or $M\leq
m_1+m_2$.
As a result, the Landau singularity is located outside the physical region, as can be seen from
Fig.~\ref{fig:singularity}. The $\Upsilon(10860)\to\omega\chi_{b0}$
process, denoted by a star in the plot, is not far from the singularity
trajectory. However, for $M=M_{\Upsilon(10860)}$, the loop function does not
diverge at the solutions of Eq.~\eqref{eq:NRsingularity} because the divergences
from both arctan functions cancel with each other in this case.

\end{appendix}

\medskip


\end{document}